\documentclass[]{article}
\usepackage{psfig,sprocl_my}
\def\be{\begin{equation}}
\def\ee{\end{equation}}
\def\bea{\begin{eqnarray}}
\def\eea{\end{eqnarray}}
\newcommand{\msbar}{{\overline{\mbox{MS}}}}
\newcommand{\xgc}{x_{\rm cut}}

\newcommand{\nn}{\nonumber}
\renewcommand{\thefootnote}{\fnsymbol{footnote}}

\begin{document}

\title{THE REACTION $e^+e^-\to t\bar{t}g$ AT NEXT-TO-LEADING ORDER
IN $\alpha_s${\footnote{Talk presented at {\it The Worldwide Study on Physics
and Experiments with Future Linear $e^+e^-$ Colliders, Sitges, Spain, April 28 - May 5 1999}.}}}

\author{ ARND BRANDENBURG }

\address{DESY Theory Group\\ D-22603 Hamburg, Germany}


\maketitle\abstracts{
The production of top quark pairs together with a hard gluon
in $e^+e^-$ annihilation is studied including  
next-to-leading order corrections in the strong coupling.}
  
\section{Motivation}
\label{sec:mot}
\setcounter{footnote}{0}
\def\thefootnote{\alph{footnote}}
\def\@makefnmark{{$\!^{\@thefnmark}$}}
The production of top quark pairs
in association with (at least) one additional parton $X$
carrying a hard momentum $k_X$,
\bea
\label{reac}
e^+(p_+) + e^-(p_-) \to t(k_t) + \bar{t}(k_{\bar{t}}) + X(k_X),
\eea
is interesting for several reasons:
A comparison of the cross section for (\ref{reac})
(which will be properly defined below) 
to the inclusive $t\bar{t}$ production cross section allows 
a measurement of the strong coupling constant $\alpha_s$.
A deviation of this measurement from the 
determination of $\alpha_s$ using event samples containing 
light quark jets would indicate a violation of the ``flavour
independence'' of the strong interactions --- i.e., would
point towards new physics phenomena connected with the top quark.
Specific examples of non-standard interactions that could
be probed by reaction (\ref{reac}) are possible anomalous
couplings of the top quark to photons, $Z$-bosons and gluons.
It has been shown \cite{Ri96} that a 
large anomalous chromomagnetic $t\bar{t}g$ coupling would 
modify the gluon energy spectrum in $e^+e^-\to t\bar{t}g$.
Furthermore, symmetry tests can be performed utilizing 
the richer kinematic structure of the final state in
(\ref{reac}). These contain tests of the CP symmetry 
\cite{BaAtEiSo96} and
the search for final state rescattering (`T'-violating) effects  
using triple momentum correlations.\cite{BrDiSh96}  
Both the search for heavy quark anomalous
couplings \cite{sld1} and the symmetry tests \cite{sld2} 
have been shown to be experimentally
feasible in the case of $b\bar{b}g$ production at the $Z$
resonance, and it will be interesting to perform similar
studies with top quarks. In order to unravel
possible deviations from QCD expectations, it is mandatory
to analyse reaction (\ref{reac}) at next-to-leading order
in $\alpha_s$.
\section{Leading order analysis}
At order $\alpha_s$, the additional parton $X$ 
in (\ref{reac}) can only be a gluon.
The cross section for $e^+e^-\to t\bar{t}g$ develops a soft
singularity as the gluon energy goes to zero. 
An infrared finite cross section can be defined
by demanding $x_g \equiv 2 E_g/\sqrt{s} > \xgc$, 
where $E_g$ is the gluon energy in the c.m. system, $\sqrt{s}$ is the
c.m. energy, and $\xgc$ is some preset number. 
Since $x_g$ is no useful variable for final states with
four or more partons (which are  relevant
at higher orders in $\alpha_s$), we use instead 
the scaled $t\bar{t}$ invariant mass square $x_{t\bar{t}}
=(k_t+k_{\bar{t}})^2/s$, i.e. replace the condition
$x_g>\xgc$ by the (at LO equivalent) condition:{\footnote{Other, at LO 
equivalent, choices for the cut condition are of course possible, 
e.g. $2E_X/\sqrt{s} >\xgc$, where $E_X=\sqrt{s}-E_t-E_{\bar{t}}$, and
all energies are defined in the c.m. system.}}
\bea
1-x_{t\bar{t}}>\xgc.
\eea
We define $r(\xgc)$ as the fraction of $t\bar{t}X$ events for which 
$1-x_{t\bar{t}}>\xgc$ with respect to all $t\bar{t}$ events,
\bea
\label{rlo}
r(\xgc) = \frac{\sigma\left(e^+e^-\to t\bar{t}X;\ 
1-x_{t\bar{t}}>\xgc\right)}
{\sigma_{\rm tot}(e^+e^-\to t\bar{t})}\equiv \frac{\sigma_3(\xgc)}
{\sigma_{\rm tot}}.
\eea
At leading order, a compact analytic result can be derived for
$r(\xgc)$.\cite{Br99}
A remark is in order concerning the experimental
distinction of $t\bar{t}$ events with a gluon radiated off the $t$
or $\bar{t}$ from events in which the gluon is radiated off
the $b$ or $\bar{b}$ produced in the decays of the top quark pairs.
It has been shown \cite{MaOr98,Or99} 
that the following two constraints
efficiently select events where the gluon is produced in association with 
the top quark pair:
\bea
\label{orr}
E_g &>& \frac{\sqrt{s}}{2}\xgc \gg \Gamma_t,  \nn \\
(m_t-2\Gamma_t)^2 &\le& (k_{W^{\pm}}+k_{b(\bar{b})})^2
\le (m_t+2\Gamma_t)^2.
\eea
By requiring that the invariant mass of the $Wb$ system lies in the
vicinity of the top quark mass, the probability that a highly
energetic gluon jet ($E_g \gg \Gamma_t \approx 1.4$ GeV) 
is emitted from the $b$ or $\bar{b}$ is very small.
\section{Results at next-to-leading order}
To evaluate the fraction $r(\xgc)$ at order $\alpha_s^2$ 
we need two ingredients:
First, the total cross section $\sigma_{\rm tot}(e^+e^-\to t\bar{t})$
to order $\alpha_s$, which is well-known.\cite{ChKuKw96}
Second, we need $\sigma_3(\xgc)$ at order $\alpha_s^2$, i.e.
both virtual and real corrections to the reaction $e^+e^-\to t\bar{t}g$.
For this we apply the results and techniques
developed for the analogous case
of three-jet production at NLO involving massive $b$ quarks.\cite{BrUw98} 
The real corrections consist of the processes
$e^+e^- \to t\bar{t}gg,\ e^+e^- \to t\bar{t}q\bar{q}\ 
(q=u,d,s,c,b)$. 
If $\sqrt{s}>4m_t$, the production of
{\it two} $t\bar{t}$ pairs becomes possible. However, these
rather spectacular events are extremely rare for the c.m. energies
considered below and contributions from 
the process $e^+e^-\to t\bar{t}t\bar{t}$
can therefore be neglected.
We renormalize the coupling in the modified minimal subtraction
($\msbar$) scheme. For the results presented in Figs. 1 and 2, the
top quark mass is defined as the perturbative pole mass.

\begin{figure}
\psfig{figure=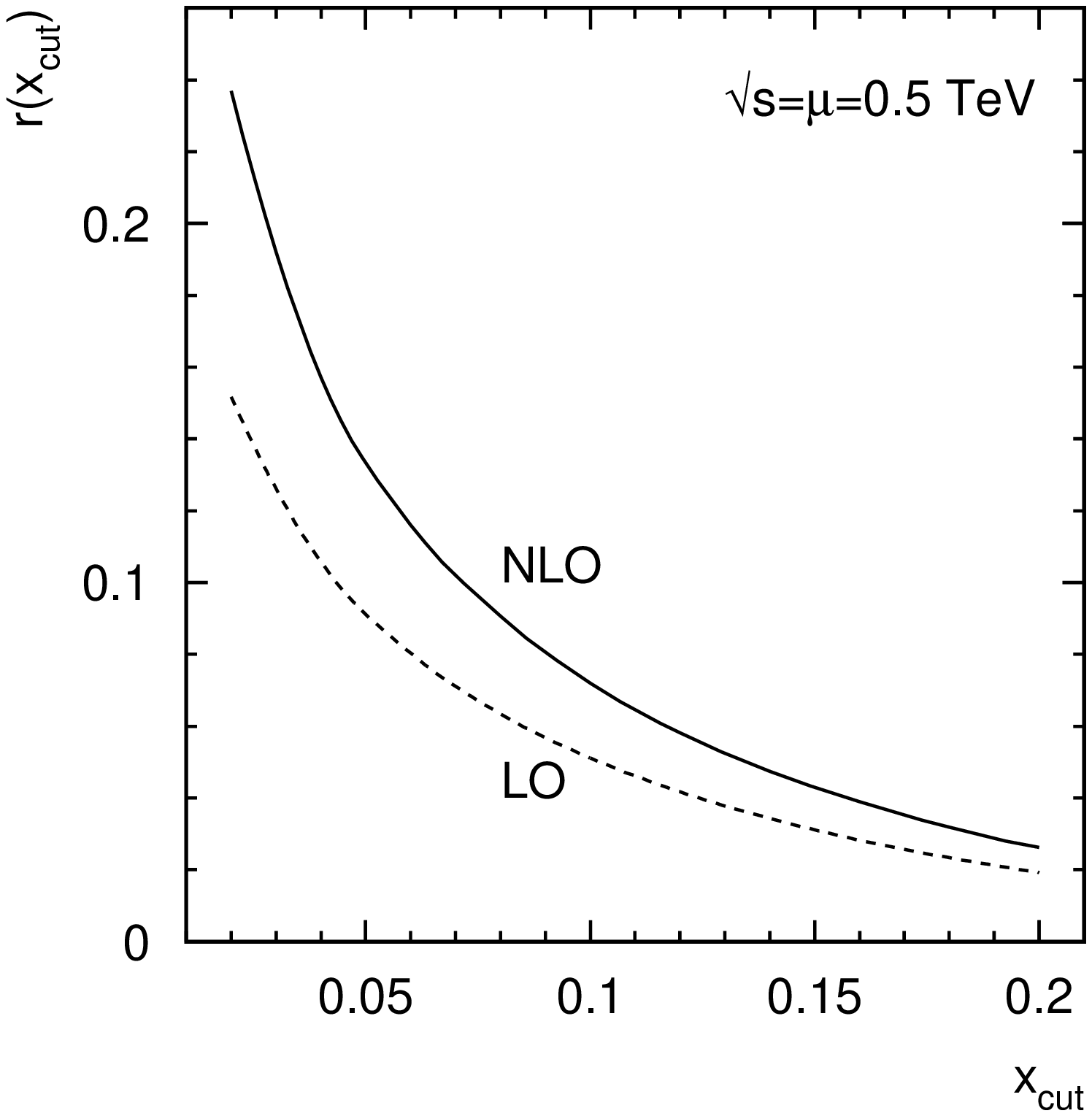,height=2.5in}
\begin{picture}(0,0)
\put(180,12){\psfig{figure=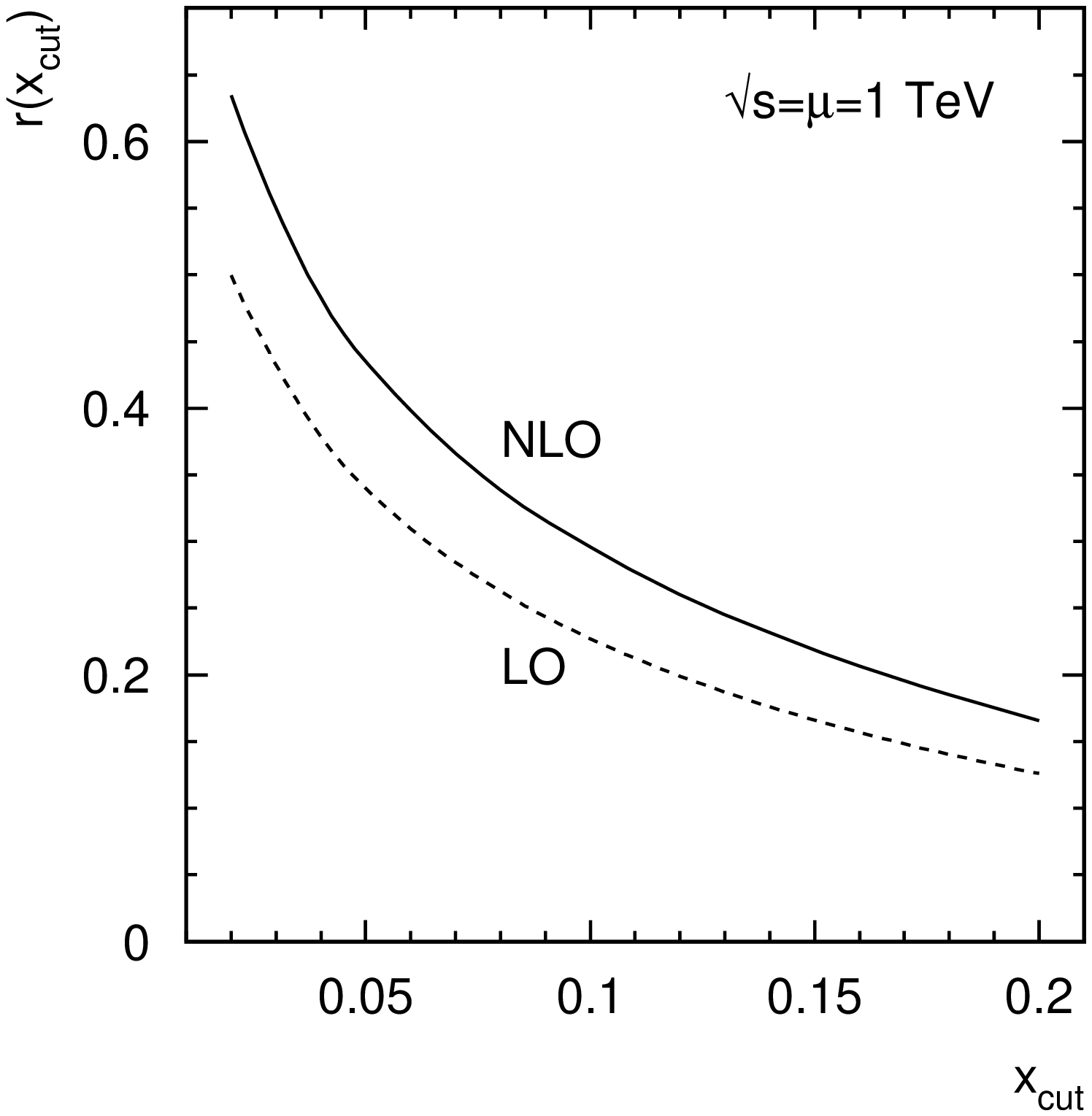,height=2.5in}}
\end{picture}
\vskip -0.75cm
\caption{Fraction $r(\xgc)$ as a function of $\xgc$ at 
$\sqrt{s}=\mu = 0.5$ and 1 TeV.
\label{fig:ycut}}
\end{figure}

Fig. 1 shows the LO and NLO results
for $r$ as a function of $\xgc$ at $\sqrt{s}=0.5$ TeV 
and $\sqrt{s}=1$ TeV. The renormalization scale is set to
$\mu=\sqrt{s}$, and for the top quark pole mass we use
$m_t=175$ GeV. (A detailed discussion of the scale
and mass renormalization scheme dependence is presented 
elsewhere.\cite{Br99}) For $\sqrt{s}=0.5$ TeV, the relative 
size of the QCD corrections 
varies between 56\% (at $\xgc=0.02$) and 36\% (at $\xgc=0.2$).
At $\sqrt{s}=1$ TeV, the QCD corrections are roughly
constant as $\xgc$ is varied and of the order of 30\%.
In Fig. 2 we plot distributions of the 
cross section $\sigma_3(\xgc)$ (defined in (\ref{rlo})) 
w.r.t. to the scaled top quark
energy $x=2E_t/\sqrt{s}$ and the variable $1-x_{t\bar{t}}$.  
The distributions are
normalized to the total $t\bar{t}$ cross section $\sigma_{\rm tot}$,
and we set $\sqrt{s}=\mu=0.5$ TeV and 
$\xgc=0.1$.
\begin{figure}[h,b]
\psfig{figure=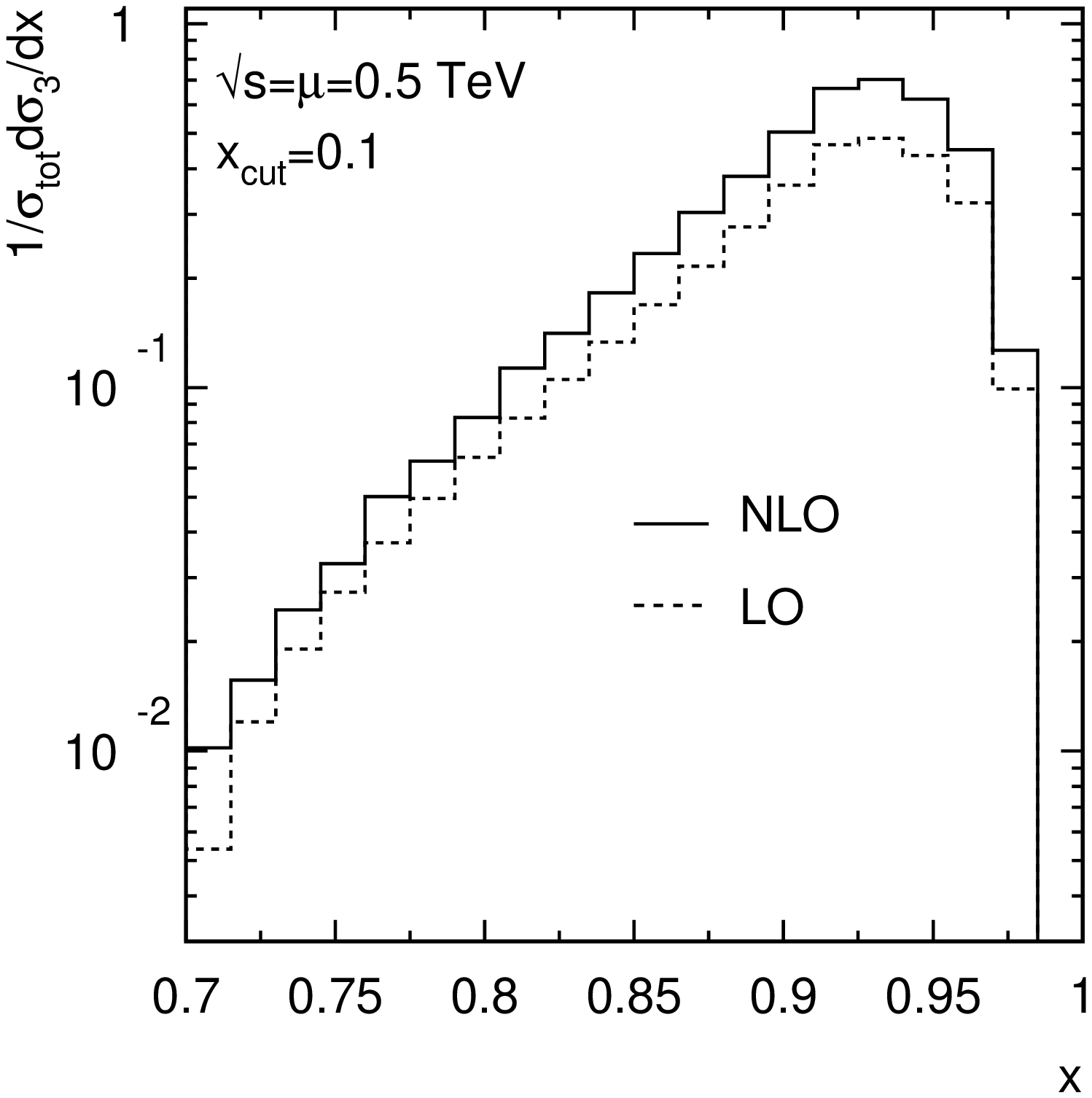,height=2.5in}
\begin{picture}(0,0)
\put(180,12){\psfig{figure=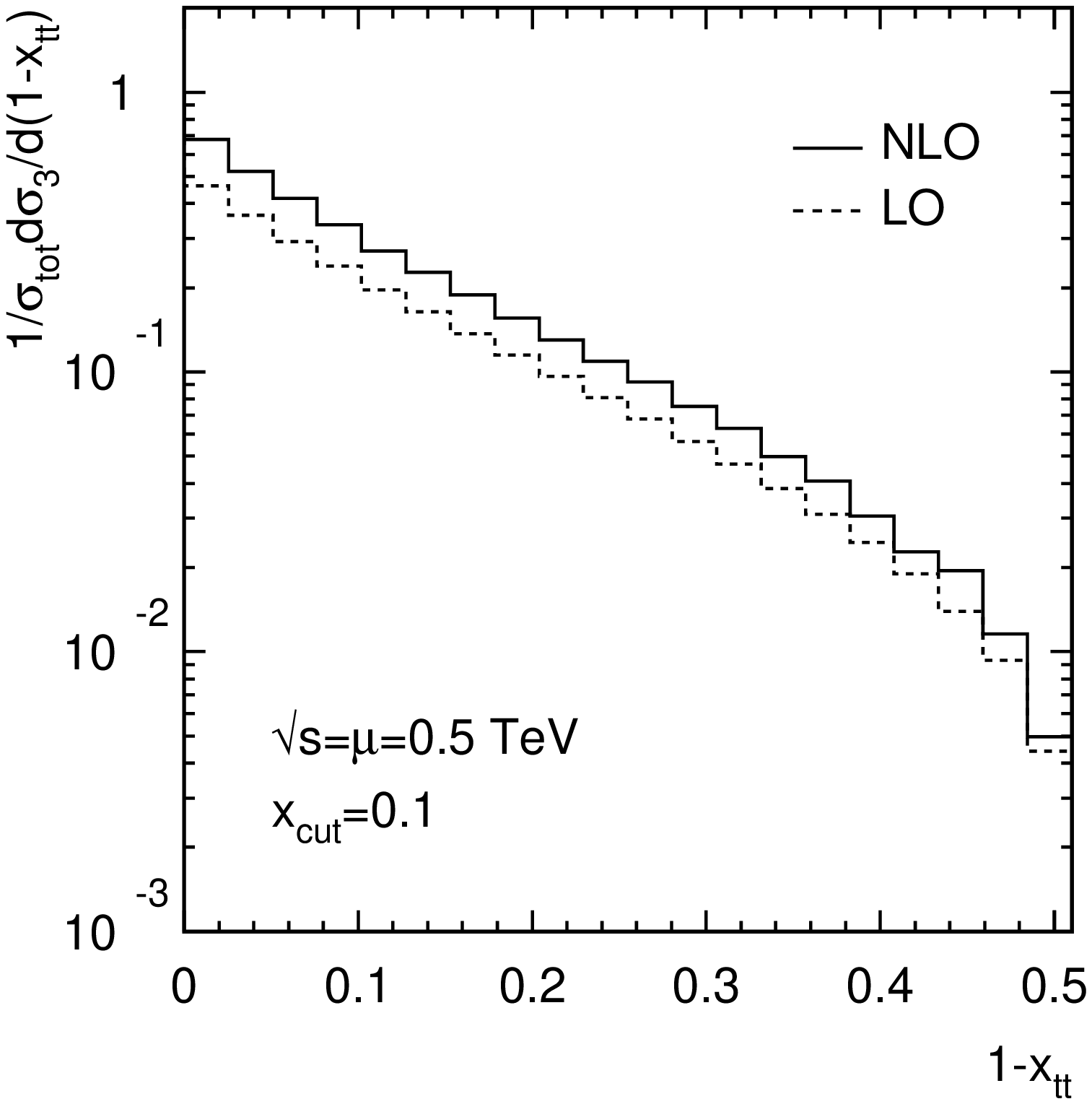,height=2.5in}}
\end{picture}
\vskip -0.75cm
\caption{Distributions $1/\sigma_{\rm tot}d\sigma_3/dx$ and
$1/\sigma_{\rm tot}d\sigma_3/d(1-x_{t\bar{t}})$.
\label{fig:dist}}
\end{figure}
\par
An intriguing question is whether a measurement of
the fraction $r$ allows a direct determination of the
value of the running ($\msbar$) mass parameter $\overline{m}_t$
of the top quark at high energies. 
(Analogously, NLO results for three-jet fractions
involving $b$ quarks \cite{BeBrUw97,RoSaBi97,BrUw98}
have been used to extract a value for $\overline{m}_b(\mu=m_Z)$
from the high-statistics LEP \cite{delphi98} and SLD 
\cite{bu99} data.) We consider here as a case study
the result for $r(\xgc=0.1)$ at $\sqrt{s}=1$ TeV. 
We first express the fraction $r$ in terms 
of the running mass $\overline{m}_t$ rather than in terms 
of the pole mass. This reduces the dependence
on the renormalization scale significantly.\cite{Br99} We then  
compute  $r(\xgc=0.1)$ at NLO for 
values of  $\overline{m}_t(\mu=1 {\mbox{ TeV}})$ 
between, say, 140 and 160 GeV. (The
value obtained from the renormalization group evolution
is $\overline{m}_t(\mu=1{\mbox{ TeV}})=148.6$ GeV.)  
The fraction $r(\xgc=0.1)$ decreases by about
4\% when the running top quark mass is changed from 140 to 
160 GeV. We find that if a measurement of $r$ 
will be possible with
an error of $\pm 1$\%, the running top quark mass
at $\mu=1$ TeV could be determined up to $\pm 5$ GeV.
A statistical error of 1\% on $r$ is realistic 
with the envisioned high luminosity of a 
future linear collider operating at $\sqrt{s}=1$ TeV.
The sensitivity of the fraction $r$ 
on $\overline{m}_t(\mu=1 {\mbox{ TeV}})$
may appear rather poor, but a direct determination of the running
top quark mass at such a high scale would provide a nice
consistency check of perturbative QCD in the following way:
The direct measurement could be compared 
with the value for $\overline{m}_t(\mu=1 {\mbox{ TeV}})$ which one
obtains by a conversion and evolution of the ``1S'' 
or ``potential subtracted'' \cite{Te99} 
mass to be extracted from the threshold scan of
$\sigma_{\rm tot}$.

In summary, the production of top quark pairs together with one or
more additional hard partons at a future $e^+e^-$ linear collider
will be an exciting new testing ground for perturbative QCD
and possible new interactions of the top quark.

\section*{Acknowledgments}
This work was supported by a Heisenberg grant of the DFG.



\section*{References}
\begin{thebibliography}{99}
\bibitem{Ri96} T.G. Rizzo, Phys. Rev. D {\bf 50}, 4478 (1994); 
SLAC-PUB-7317, hep-ph/9610373 (1996).
\bibitem{BaAtEiSo96} S. Bar-Shalom, D. Atwood, G. Eilam, and A. Soni,
Z. Phys. C {\bf 72}, 79 (1996).
\bibitem{BrDiSh96} A. Brandenburg, L. Dixon, and Y. Shadmi,
Phys. Rev. D {\bf 53}, 1264 (1996). 
\bibitem{sld1} K. Abe {\it et al.} (SLD Collab.),  SLAC-PUB-7920,
hep-ex/9903004 (1999), to appear in Phys. Rev. D.
\bibitem{sld2} K. Abe {\it et al.} (SLD Collab.), Phys. Rev. Lett.
{\bf 75}, 4173 (1995); SLAC-PUB-7823 (1998).
\bibitem{Br99} A. Brandenburg, hep-ph/9904251 (1999), 
to appear in Eur. Phys. J. C.
\bibitem{MaOr98} C. Macesanu, L.H. Orr, UR-1542, ER/40685/921, 
hep-ph/9808403 (1998).
\bibitem{Or99} L.H. Orr, these proceedings.
\bibitem{ChKuKw96} K.G. Chetyrkin, J.H. K\"uhn, and 
A. Kwiatkowski, Phys. Rep. {\bf 277}, 189 (1996).
\bibitem{BrUw98} A. Brandenburg and P. Uwer, 
Nucl. Phys. B {\bf 515}, 279 (1998).
\bibitem{BeBrUw97} W. Bernreuther, A. Brandenburg, and P. Uwer, 
Phys. Rev. Lett. {\bf 79}, 189 (1997).
\bibitem{RoSaBi97} G. Rodrigo, A. Santamaria, and M. Bilenky,
Phys. Rev. Lett. {\bf 79}, 193 (1997).
\bibitem{delphi98} P. Abreu {\it et al.} (DELPHI Collab.),
Phys. Lett. {\bf B 418}, 430 (1998).
\bibitem{bu99} A. Brandenburg {\it et al.} SLAC-PUB-7915,
hep-ph/9905495 (1999).
\bibitem{Te99} T. Teubner, these proceedings; A. Hoang, these proceedings.
\end{thebibliography}
\end{document}